# Improvement in RUP Project Management via Service Monitoring: Best Practice of SOA

Sheikh Muhammad Saqib[1], Shakeel Ahmad[1], Shahid Hussain[2], Bashir Ahmad[1] and Arjamand Bano[3]

[1]Institute of Computing and Information Technology Gomal University, D.I.Khan, Pakistan
[2] Namal University, Mianwali , Pakistan
[3]Mathematics Department Gomal University, D.I.Khan, Pakistan

**Abstract--** Management of project planning, monitoring, scheduling, estimation and risk management are critical issues faced by a project manager during development life cycle of software. In RUP, project management is considered as core discipline whose activities are carried in all phases during development of software products. On other side service monitoring is considered as best practice of SOA which leads to availability, auditing, debugging and tracing process. In this paper, authors define a strategy to incorporate the service monitoring of SOA into RUP to improve the artifacts of project management activities. Moreover, the authors define the rules to implement the features of service monitoring, which help the project manager to carry on activities in well define manner. Proposed frame work is implemented on RB (Resuming Bank) application and obtained improved results on PM (Project Management) work.

## 1. INTRODUCTION

Software projects which are developed through RUP process model are solution oriented and object oriented. It provides a disciplined approach to assigning tasks and responsibilities within a development organization [3]. The perfect planning and management helps to carried out activities smoothly in order to achieve objectives in specific period of time. This will be only possible when all is done within the spectrum of project management. In different process model, special emphasis is given to project management activities such as in Rational Unified Process (RUP) model. In RUP, project management is considered as core discipline whose activities are carried out through out all phases. Software project management is the skill of balancing competing objectives, planning of resources, monitoring the status of work, managing risk, and overcoming constraints to deliver a product that can satisfy the customer and end users [3][4].

Service Oriented Architecture (SOA) is a set of rules, practices and architecture of business services to business practices. It is a logical way to design a software concept that provides services to either end users application or other services distributed over a network.[1]. The reasons for popularity of SOA are its best practices. Among these best practices of SOA, service monitoring is that one in which typical calculated SOA planning effort is needed. Service Monitoring helps to improve the availability, auditing, debugging and tracing process relevant to project manager in different phases of RUP.

## 2. RATIONAL UNIFIED PROCESS (RUP)

RUP is a software engineering process model, which provides a disciplined approach to assigning tasks and responsibilities within a development environment. The goal of RUP is to produce high quality software that meets the needs of its end users within a predictable schedule and budget [5]. The RUP development process consists of four phases and number of iterations, where each one of them has the purpose of producing a demonstrable part of the software project. The RUP's Phases are inception, elaboration, construction and transition. In inception phase of RUP the main focus is given to needs of stakeholder via defining the critical use cases and candidate architectures.  In elaboration phase of RUP, the candidate architectures are elaborated and completely described, use case and actors are also identified and described and finally a prototype is created. In construction phase of RUP, all the remaining components and applications are developed, integrated into the product and are tested. In transition phase of RUP, product is released with user acceptance, modification or addition of extra features takes place and manuals are written for users.

In RUP development process is supported by number of best practices such as develop software iteratively, manage requirements,  use component based architectures, visually model software, verify software quality and control changes to software. As a summary, RUP consists of a robust methodology, were small iterations, elaborate modeling sessions, constant testing and a common design language can be the better solution for creating software applications of great quality, error free and that meet all the costumer requirements [6].

### 3. SERVICE ORIENTED ARCHITECTURE (SOA)

Service Oriented Architecture can allow a business environment with a flexible communications and processing environment. SOA achieves this by managing independent, reusable automated business process and systems functions as services and providing a healthy and safe foundation for leveraging these services. It introduces logical service based architecture to create a reactive and adoptive environment [1]. SOA is of high importance especially for businesses, because the increasing importance of services in the economy of developed countries. Enterprise Architecture (EA) is a planning, governance, and innovation function. SOA is one philosophy or framework within enterprise architecture (EA), the goals and objectives of which are alignment with the business and business goals, which include [6][7].

- Reducing overall total cost of ownership (TCO)
- Improving time to market
- Achieving business agility
- Fostering innovation
- Enabling compliance
- Improving the top and / or bottom line
- Increasing customer satisfaction and retention

Further, SOA is increasingly a significant part of EA and has implications for key aspects of EA such as business architecture, application architecture, information architecture and technical architecture.

### 4. SOA PRACTICES

Like other methodologies, SOA also have best practices which help in certain key areas of SOA applications such as in SOA planning effort. The main key areas for best practices of SOA are [8].

- Service Monitoring
- Exception Management
- Version Management
- Service Management and Deployment
- Policy and Security Considerations
- Service Level Agreements
- Service Directory

  In proposed strategy, authors consider the service monitoring key area of SOA with its best practices and incorporate these into RUP for improvement in project management process.

### 5. SERVICE MONITORING

Service monitoring is the key area of SOA application which helps to audit, debug and traceability of software applications The main best practices in SM (Service Monitoring) area of SOA are:

*Availability (Av):*
The active availability of a service is generally defined within its accompanying service level agreement (SLA). Common availability considerations and guarantees include:
When will the service be active and available?
Will all dependencies for this service also be available during the scheduling of Service availability?

*Auditing (Au):*
Auditing is the analysis and reporting of logged information. The data collected at this stage should answer questions such as:
How exactly is a service being used?
Who is getting the benefits from service?
What do the general request and response messages look like?

*Performance and Metrics (PM):*
Performance metrics and statistics for each Web service are kept so that services can be monitored to avoid potential runtime issues, such as performance bottlenecks and fault counts. It is often required and recommended that automated notification mechanisms be attached to collected statistics so that action can be taken well before an infrastructure's limitations are tested.

*Debugging and Tracing (DT):*
The ability to optimize and track service activity during development and after deployment is important, especially for maintenance purposes. This focuses more on the service hosting platform's skill to provide front-end tools that allow for targeted exploration of specific activities (or sub-activities) as opposed to common general reporting features.

### 6. ARTIFACTS OF PROJECT MANAGEMENT DISCIPLINE

In RUP the activities of project manager are perform through the whole development life cycle and there are number of artifacts which are concerned with activities of project manager and needs to tailor [3]. The list of artifacts, its purpose and recommendation is shown in Table-1.

TABLE - 1. ARTIFACTS OF PROJECT MANAGEMENT

| Artifact | Purpose | Tailoring (Optional, Recommended) |
| --- | --- | --- |
| Business Case | Used to determine whether or not the project is worth investing in. | Recommended. |
| Iteration Assessment | Captures the effect of an iteration, the degree to which the evaluation criteria were met, lessons learned, and changes to be done. | Recommended. |
| Iteration Plan | The detailed plan for the iteration, including the time-sequence of tasks and resources. | Recommended. |
| Software Development Plan Measurement Plan Problem Resolution Plan Product Acceptance Plan Quality Assurance Plan Risk Management Plan | Includes all information required to manage the project. | All projects need some planning in order to manage a project. Smaller, less complex projects, may have a single document capturing the project plan. Larger, more complex, or more formal projects will require multiple separate sub plans. |
| Project Measurements | This is the repository of all measurements related to the project. | Recommended for most projects. On many projects, only a few measures are used, such as cost and progress measures. A metrics database is required only when there is large amount of metrics data to be managed. Many organizations gather metrics data from multiple projects in order to glean information to apply to future projects. |
| Review Record | Captures the results of a review of one or more project artifacts. Review records can avoid misunderstandings of decisions made during a review. They also serve as evidence to stakeholders that project artifacts are being reviewed. | Recommended for most projects. Most projects will want to record decisions made in meetings with the customer and other key stakeholders, in order to ensure a common understanding. Reviews records for other reviews may or may not be formally captured, depending on the review formality applied by the particular project. |
| Risk List | This is a prioritized list of project risks. | Recommended. May be just a section in the Software Development Plan. |
| Status Assessment | Used to capture a snapshot of project status, including progress, management issues, technical issues, and risks. | Recommended. The Status Assessment may be shared with the Iteration Assessment if the iterations are frequent (one each month). If iterations are long, there will be a need for middle Status Assessments. |
| Work Order | This is a negotiated agreement between the Project Manager and the staff to perform a particular activity, or set of activities, under a defined schedule and with certain deliverables, effort, and resource constraints. | Recommended for most projects. May be implemented using Change Requests. |

## 7. INCORPORATING BEST PRACTICES OF SERVICE MONITORING WITH ACTIVITIES AND ARTIFACTS OF PROJECT MANAGEMENT IN RUP

In each phase of RUP, project manager perform number of activities relevant to project management and generate artifacts. According to proposed strategy best practices of service monitoring are mapped into activities/artifacts of project manager. This mapping process is shown for each phase.

*Inception and Elaboration Phase:*

Inception phase is the core phase of RUP and here any wrong decisions about gathering business cases would lead to failure of project. The idea of a business case is to capture the logic for initiating a project or task. Incomplete business cases cause the problem for integration of development case. This problem may be solved with the availability best practice of service monitoring which shows that when service will be active and available. The conflict in collaboration of different roles/stakeholders can be overcome due to auditing and availability best practices of service monitoring. These best practices would helps to show the clear description of each role/stakeholder, its work and the time when work will be done. Moreover, auditing best practice can be mapped into iteration plans and its assessment. Besides these auditing and availability best practices of service monitoring can be included in risk management life cycle such as availability is risk mitigation process. Similarly assessment status and artifacts and corresponding mapped best practices of SM.

project measurement can be enhanced by incorporating the performance and matrices, debugging and auditing best practice of service monitoring. Table-2 shows the list of

| PROJECT MANAGEMENT ACTIVITIES | DRAWBACKS | USING PRACTICE OF SERVICE MONITORING | RESULTS |
|---|---|---|---|
| Business Case | Integration of Development Case | Availability | No chance of missed activity. |
| Software Development Plan | Complex project 's interaction with many stockholders | Auditing, Availability. | All Stack holders can communicate with each others |
| Iteration Plan | To remove the flaws of other artifacts which depends on iteration plan | Auditing. | Iteration plan become clear and fine after auditing |
| Iteration Assessment | As These are not updated. And can be blocked due to wrong iteration plan | No Need of Service Monitoring Practice. | Auditable iteration plan, No flaws in iteration assessment. |
| Status Assessment | Need of intermediate Status Assessment | Performance metrics | No need of intermediate Status Assessment |
| Project Measurements. | Need of automated software data collection agent | When status Assessment is created on the basis of Performance metrics then there is no need of automated software data collection agent. | No need of automated software data collection agent |

TABLE-2. APPLYING BEST PRACTICES OF SERVICE MONITORING WITH ARITFACTS OF INCEPTION AND ELEBORATION PHASE

*Construction Phase:*

In this phase, the best practices of service monitoring such as auditing, debugging and performance metrics are incorporated within the activities of project manager and it would leads to enhance the artifacts of project management activities. Construction phase is started at the end of elaboration phase. Here in this phase, activities of project management discipline will be still continued. This phase of RUP shows the development process of software so here performance matrices would be included in designing process and auditing and debugging would be include in coding process of this phase. SQA members, internal developer or any third party member can do auditing and debugging process to check the functionality of software.

*Transition Phase:*

In this phase, the best practices of service monitoring such as auditing, debugging and tracing are incorporated within the activities of project manager and it would leads to enhance the artifacts of project management activities. Due to this the artifacts of project manager relevant to testing and deployment process will be enhanced.

The mapping process of best practices of service monitoring into the activities of project manager in each phase is shown in Fig. 1.

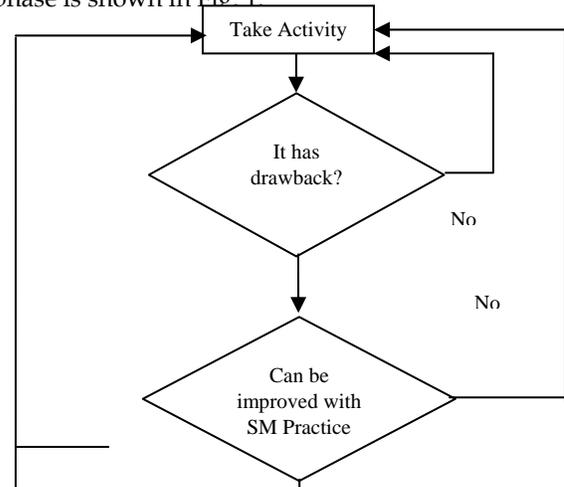

Fig-1. Flowchart for improvement

**RB** (Resuming Bank) is student level project based on web and desktop application which is developed through RUP. Its major users are Administrator, Evaluator, and Applier. Administrator can set Rules scenery, job status (Online or Off line), start and last date for application etc. Evaluator can neglect the applier if he is not qualified. Applier can set his online resume which will be stored in resuming bank through web interface and now he can apply online. SM (Service Monitoring) is applied on PM( Project Management) work of RUP with some activities of RB.

Applying proposed methodology on RB, we can improve project manager's work.

Table-3 shows that when SM practices on few activities of RB in PM work are applied, then some needs/ requirements like users(who will use the activity), interfaces(application interface either desktop or web) are identified. The letter I, E, C and T in RUP phase's column of Table-3 represent the Inception, Elaboration, Construction and Transition for each activity.

TABLE-3. APPLYING SM PRACTICES ON RB ACTIVITIES FOR PM DISCIPLINE OF RUP

| Activity No | ORB Activities | Requirements of RB Determines from SM on PM | | | RUP Phases | Practice of SOA (SM) | | | |
|---|---|---|---|---|---|---|---|---|---|
| | | Users | Interface | Desktop Web | | Av | Au | PM | DT |
| 1 | Rules Setting | Administrator | YES | | I | Y | Y | Y | N |
| | | | | | E | Y | Y | Y | N |
| | | | | | C | N | Y | Y | Y |
| | | | | | T | N | Y | N | Y |
| | | | NO | | I | Y | Y | Y | N |
| | | | | | E | Y | Y | Y | N |
| | | | | | C | N | Y | Y | Y |
| | | | | | T | N | Y | N | Y |
| 2 | Job Status | Administrator | YES | | I | Y | Y | Y | N |
| | | | | | E | Y | Y | Y | N |
| | | | | | C | N | Y | Y | Y |
| | | | | | T | N | Y | N | Y |
| | | | YES | | I | Y | Y | Y | N |
| | | | | | E | Y | Y | Y | N |
| | | | | | C | N | Y | Y | Y |
| | | | | | T | N | Y | N | Y |
| 3 | Scrutiny | Evaluator | YES | | I | Y | Y | Y | N |
| | | | | | E | Y | Y | Y | N |
| | | | | | C | N | Y | Y | Y |
| | | | | | T | N | Y | N | Y |
| | | | NO | | I | Y | Y | Y | N |
| | | | | | E | Y | Y | Y | N |
| | | | | | C | N | Y | Y | Y |
| | | | | | T | N | Y | N | Y |
| 4 | Online Resume Setting | Applier | NO | | I | Y | Y | Y | N |
| | | | | | E | Y | Y | Y | N |
| | | | | | C | N | Y | Y | Y |
| | | | | | T | N | Y | N | Y |
| | | | YES | | I | Y | Y | Y | N |
| | | | | | E | Y | Y | Y | N |
| | | | | | C | N | Y | Y | Y |
| | | | | | T | N | Y | N | Y |

## 8. Conclusion

There are number of factors for a project manager of an organization which reflects the delivery of product on proper time such as knowledge, complete understanding of scope and resources and monitoring process of work. In RUP, project manager do a lot of work to complete a successful project so that's why project management is considered as separate discipline. On other side service monitoring is the key area of SOA applications with number of best practices such as availability, auditing, tracing, debugging and performance matrices. All these best practices help to improve the monitoring process of a project. In this paper authors plan a strategy which guides the project manager for incorporating the best practices of service monitoring into concern activities of project manager in each phase. Proposed methodology will enhance all project management artifacts of project manager; we have implemented it on some activities of RB application and determine their some requirements which can make PM work of RUP efficiently.

**Mr. Sheikh Muhammad Saqib** is an MS student in Institute of Computing and information technology, Gomal University D.I.Khan, Pakistan. He has got distinction throughout his academic carrier. He is doing specialization in the area of RUP model and SOA. He has published a research paper in international journal on mapping of SOA and RUP.

**Dr. Shakeel Ahmad** Dr. Shakeel Ahmad received his B.Sc. with distinction from Gomal University, Pakistan (1986) and M.Sc. (Computer Science) from Qauid-e-Azam University, Pakistan (1990). He served for 10 years as a lecturer in Institute of Computing and Information Technology (ICIT), Gomal University Pakistan.
Now he is serving as an Assistant Professor in ICIT, Gomal University Pakistan since 2001. He is among a senior faculty member of ICIT. Mr. Shakeel Ahmad received his PhD degree (2007) in Performance Analysis of Finite Capacity Queue under Complex Buffer Management Scheme.
Mr. Shakeel's research has mainly focused on developing cost effective analytical models for measuring the performance of complex queuing networks with finite capacities. His research interest includes Performance modeling, Optimization of congestion control techniques, Software Engineering, Software Refactoring, Network security, Routing Protocols and Electronic learning. He has produced many publications in Journal of international repute and also presented papers in International conferences.

**Mr. Shahid Hussain** has done MS in Software Engineering from City University, Peshawar, Pakistan. He has got distinction throughout his academic carrier. He has done his research by introducing best practices in different software process models. I have introduced a new role communication model in RUP using pairing programming as best practice.. Recently, I am working as course chair cum Lecturer in Namal College, an associate college of University of Bradford. Moreover, I have published much research paper in different national/international journals and conferences such as MySec04, JDCTA, IJCSIS, NCICT, ZABIST.